\documentclass[aps,prb,showpacs,twocolumn,floats]{revtex4-1}
\usepackage{amssymb}
\usepackage{amsbsy}
\usepackage{amsmath}
\usepackage{epsfig}
\usepackage{amsfonts}
\usepackage{graphicx}
\usepackage{amssymb}
\usepackage{upgreek}
\usepackage{dsfont}
\usepackage{bm}
\usepackage{color}
\usepackage{hyperref}
\usepackage[tight]{subfigure}
\begin{document}

\title {Interplay between topology and disorder in a two-dimensional semi-Dirac material}
\author{P. V. Sriluckshmy, Kush Saha, Roderich Moessner}
\affiliation{Max-Planck Institute for the Physics of Complex Systems\\
Noethnitzer Strasse 38, 01187 Dresden, Germany }
\date{\today}
\begin{abstract}
We investigate the role of disorder in a two-dimensional semi-Dirac material characterized by a linear dispersion in one direction and a parabolic dispersion in the orthogonal direction. Using the self-consistent Born approximation, we show that disorder can drive a topological Lifshitz transition from an insulator to a semi metal, as it generates a momentum- independent off-diagonal contribution to the self-energy. Breaking time-reversal symmetry enriches the topological phase diagram with three distinct regimes- single-node trivial, two-node trivial, and two-node Chern. We find that disorder can drive topological transitions from both the single- and two-node trivial to the two-node Chern regime. We further analyze these transitions in an appropriate tight-binding Hamiltonian of an anisotropic hexagonal lattice by calculating the real-space Chern number. Additionally, we compute the disorder-averaged entanglement entropy  which signals  both the topological Lifshitz and Chern transition as a function of the anisotropy of the hexagonal lattice. Finally, we discuss experimental aspects of our results. 
\end{abstract}

\maketitle
\section{ Introduction} 
Since the discovery of graphene, two-dimensional (2D) Dirac materials continue to emerge as an important and promising field of research in condensed matter physics. The presence of gapless Dirac nodes in these materials has lead to many exotic electronic properties \cite{castro}, the most striking feature being the emergence of topological states in the absence of time-reversal symmetry \cite{haldane}. The possibility of having such topological states has triggered an enormous amount of interest in searching for new materials with Dirac dispersion. Among recent proposals for 2D Dirac materials such as borophene \cite{borophene}, stanene \cite{stanene}, and silicene \cite{silicene}, the 2D semi-Dirac (SD) systems\cite{footnote1} seem to have many exotic and unusual properties \cite{banerjee,nagaosa} due to their anisotropic band dispersion: linear in one direction and parabolic along the perpendicular direction. Promising candidates for such semi-Dirac systems include TiO$_2$/V$_2$O$_3$ layered structures \cite{padro}, deformed graphene \cite{montam}, BEDT-TTF$_2$I$_3$ salt under pressure \cite{kata}, hexagonal and square lattices in the presence of magnetic field \cite{diet,del}, photonic systems \cite{yu}, etc. However, the only experimental realization for such dispersion has thus far been observed in optical lattices \cite{tilman}. The unprecedented controllability of this optical system allows one to verify different exotic properties of semi-Dirac, systems including the effect of disorder.

While the role of disorder in three-dimensional Weyl metals and 2D quantum spin Hall insulators (QSH) \cite{weyl,li,benakker,jiang,tami,roy,prodan} are well studied, the role of disorder in 2D semi-Dirac systems has received little attention \cite{carpent,peng}. Specifically, the interplay between topological states and disorder in SD systems is yet to be explored. Here, we study the effect of disorder on all different phases of a 2D semi-Dirac system \cite{footnote1}. Using the self-consistent Born approximation (SCBA), we first show that disorder can drive topological Lifshitz transition from a {\it gapless} (single-node SN)  or gapped semi-Dirac phase to a semi metallic semi-Dirac phase with {\it two} nodes (TN) of Dirac type. This is a consequence of the electronic self-energy containing a finite off-diagonal part, a feature absent in typical topological insulators \cite{benakker} or isotropic Dirac  \cite{paraj,song} systems. Furthermore, we show that breaking time-reversal symmetry in a SD system results in three distinct topological regimes: single-node trivial, two-node trivial, and two-node Chern. We show that disorder can drive a topological transition not only within a two-node regime from trivial insulating phase $(C=0)$  to a topological insulating phase ($C=1$), but it can also drive a transition from a single-node trivial to a two-node Chern regime.  The single-node trivial to two-node Chern transition involves an off-diagonal self-energy contribution that shifts the ideal semi-Dirac point in the absence of time-reversal symmetry. We analyze these topological transitions in an anisotropic honeycomb lattice model,  which is known to host semi-Dirac dispersion \cite{montam}, using a real-space Chern number. We show that entanglement entropy can serve as a tool to probe different transitions in this system. In particular, we find that the derivative of entanglement entropy shows multiple features associated with both the topological Chern and the Lifshitz transitions in the clean limit. Moreover, we show that such features survive even for weak disorder, revealing the stability of topological transitions. However, for stronger disorder the peaks diminish, leading to trivial Anderson insulators.   

\section{Model and Phases}
\begin{figure}
\includegraphics[width=0.99\linewidth]{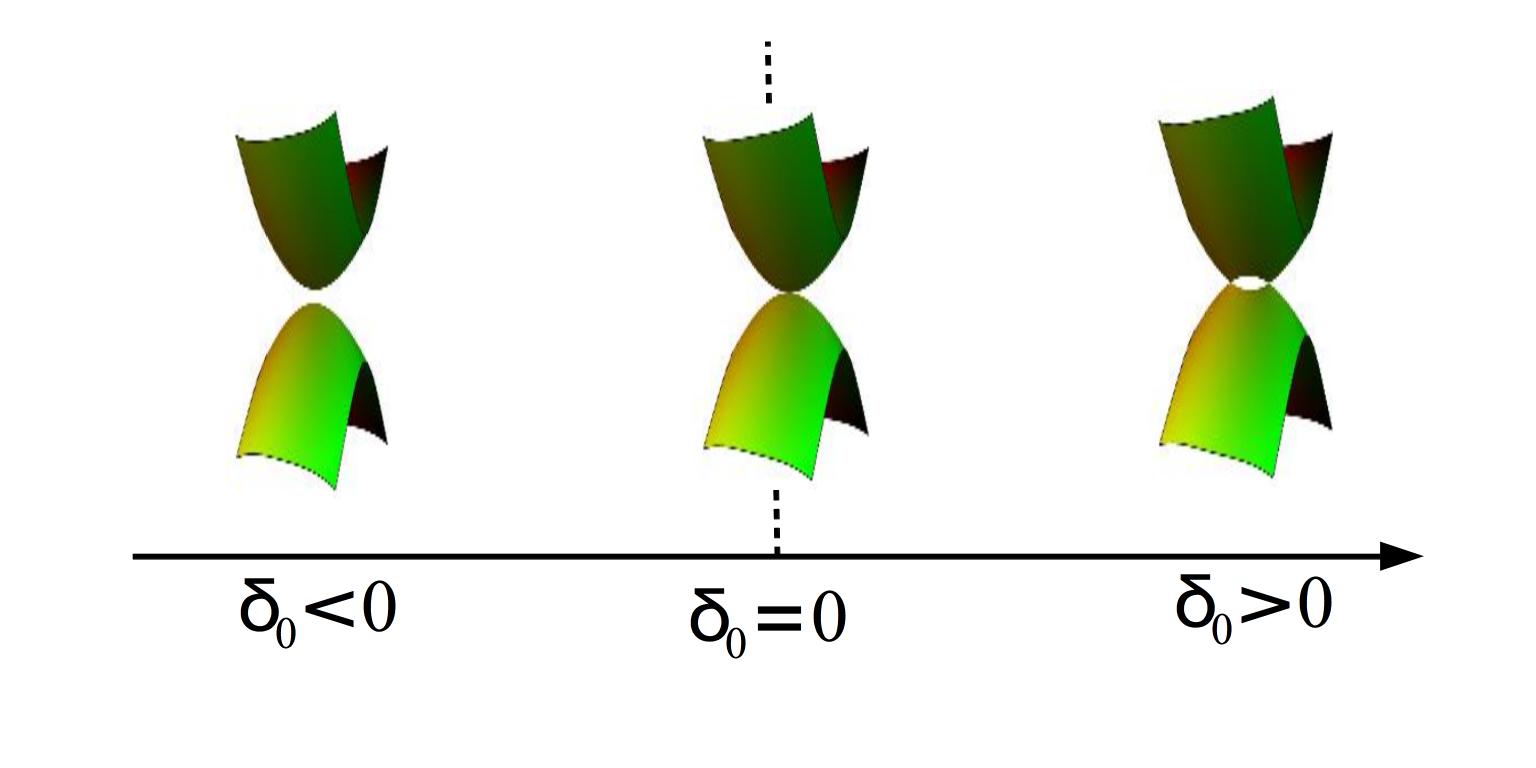}
\caption{Energy dispersion of a two-dimensional semi-Dirac spectrum [Eq.~(\ref{eq:eval})] for different values of parameter $\delta_0$. It evidences a 
topological Lifshitz transition as a function of $\delta_0$. $\delta_0 = 0$ separates the two-node and the single-node regimes.} 
\label{fig:dispersion}
\end{figure}

The low-energy model Hamiltonian describing electronic bands of a two-dimensional semi-Dirac material is \cite{del,banerjee,montam}
\begin{align}
H_{\rm SD}(\bf k)={\boldsymbol \sigma}\cdot{\bf h(\bf k)} ,
\label{ham0}
\end{align}
where ${\boldsymbol \sigma}=(\sigma_x,\sigma_y,\sigma_z)$ are the Pauli matrices, $ {\bf h (\bf k)}=(\frac{\hbar^2  k_x^2}{2\mu}-\delta_0,\hbar v_F k_y,0)$,
where ${\bf k}=(k_x, k_y)$ is the crystal momentum,  $\mu$ is the quasiparticle mass along $x$, $v_F$ is the Dirac velocity along $y$, and $\delta_0$ is the gap parameter. The energy eigenvalues are given by
\begin{align}
E^{\pm}_{k_x,k_y}=\pm \sqrt{\left(\frac{\hbar^2 k_x^2}{2\mu}-\delta_0\right)^2+\hbar^2v_F^2 k_y^2},
\label{eq:eval}
\end{align}
where $\pm$ denotes the conduction and valence band, respectively. It is well known that the variation of $\delta_0$ gives rise to three distinct phases as shown in Fig.~\ref{fig:dispersion}.  For $\delta_0=0$, the spectrum is gapless with semi-Dirac dispersion. $\delta_0<0$ corresponds to a gapped trivial insulating phase with a single node, while $\delta_0>0$ corresponds to a semi-metallic phase with two gapless Dirac nodes at $(\pm \sqrt{2\mu\delta_0/\hbar^2},0)$. Thus, $\delta_0$ plays a key role in changing the Fermi surface topology via a Lifshitz transition.

Consider a momentum-dependent perturbation to Eq. \eqref{ham0} of the form
\begin{align}
\delta H({\bf k}) = (m + \beta k_x)\sigma_z,
\end{align} 
where $m \sigma_z$ breaks particle-hole symmetry $(P=\sigma_y)$, while $\beta k_x\sigma_z$ breaks time-reversal symmetry ($\Theta=\mathcal {K}$, where $\mathcal{K}$ is the complex conjugation operator). Breaking inversion symmetry opens a gap in the semi-metallic phase with two Dirac nodes ($\delta_0>0$), and the system becomes a trivial insulator with topological invariant $C=0$. On the other hand, breaking time reversal symmetry in the same phase gives rise to topological state with $C=1$. Thus, for $\delta_0>0$, broken time-reversal and inversion symmetry lead to two distinct topological states with $C=0$ and $C=1$, respectively.  In contrast, for $\delta_0 < 0$, the system remains insulating, despite breaking the above-mentioned symmetries. We delineate the phases and label them single-node trivial (SNT), two-node trivial (TNT) and two-node Chern (TNC). The nomenclature ``single'' and ``two-node'' follows from the symmetric system where $\delta H({\bf k}) = 0$. 

\begin{figure}
\includegraphics[width=0.99\linewidth]{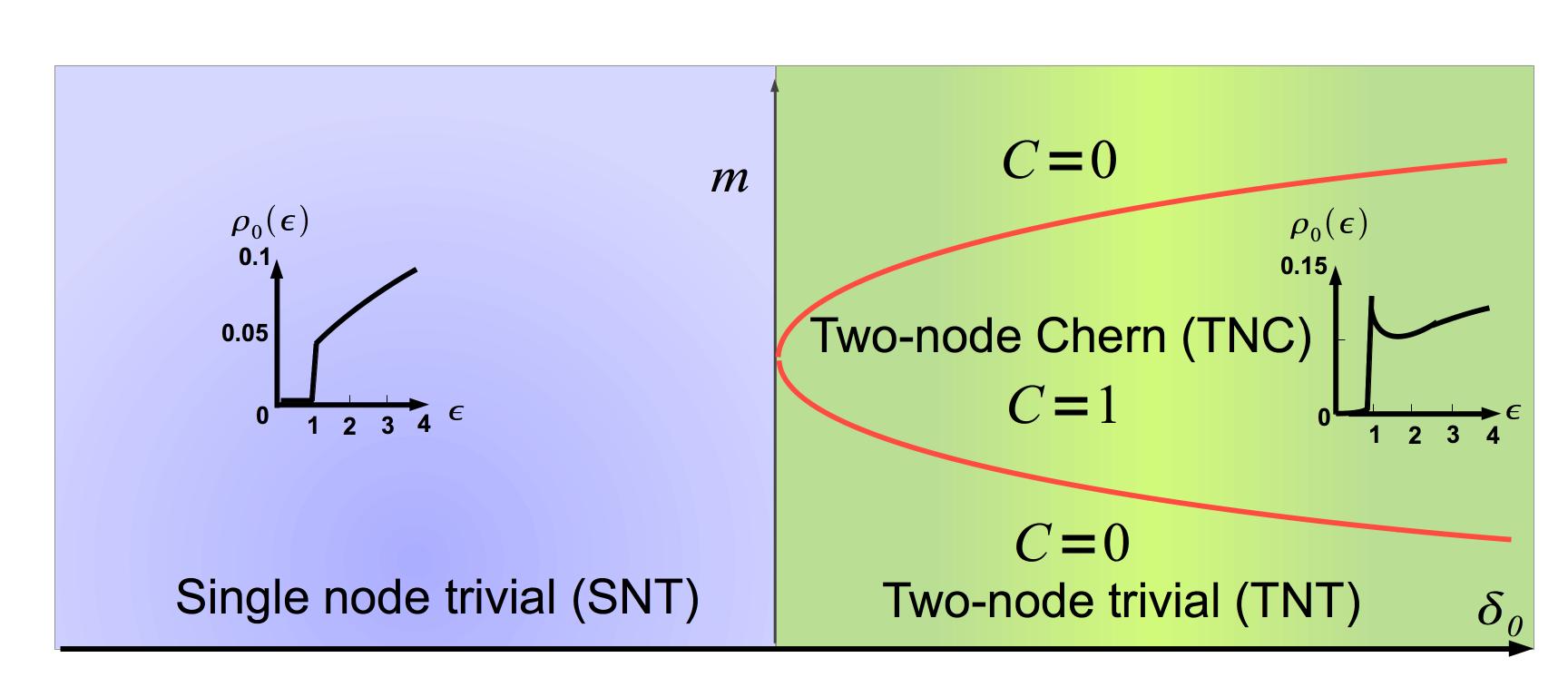}
\caption{Topological phase diagram in $\delta_0-m$ phase space for fixed $\beta$.
The phase boundary (red curve) is given by $m=\pm\beta\sqrt{2\mu\delta_0/\hbar^2}$. The two-node regime corresponds to $\delta_0>0$, while the single-node regime corresponds to $\delta_0<0$. Insets show density of states in the two regimes for $m=0$, $\beta\ne0$. Note that here we take $\hbar^2/2\mu=1$ and $\beta=1$ for simplicity.} 
\label{fig:dispersion1}
\end{figure}

Figure~\ref{fig:dispersion1} illustrates the phase diagram. For $\delta_0>0$, the phase boundary, $m=\pm \beta\sqrt{2\mu\delta_0/\hbar^2}$, separates the Chern insulating phase ($C=1$) from the trivial insulating phase ($C=0$). We also plot the density of states (DOS) in the inset of Fig.~\ref{fig:dispersion1}, which depicts the underlying topology of the Fermi surface. The kink in the DOS of the TN regime is a manifestation of the Van Hove singularity typical of Dirac-like systems. In the following, we introduce disorder in the system and discuss how it modifies the phase diagram and the topological phases.

\section{Self-consistent Born approximation}
To investigate the effect of disorder, we consider a random on-site disorder potential $U({\bf r})$, 
distributed uniformly over the interval $[-W\,,\,W]$ with  $\langle U({\bf r})U({\bf r'})\rangle=\frac{W^2}{3} \delta({\bf r-r'} )$. 
In the presence of disorder, the Green's function of the electron obeys \cite{adrouger} 
\begin{align}
G(\omega, {\bf k})=\left(\omega-H_{\rm SD}({\bf k}) - \delta H({\bf k}) -\Sigma\right)^{-1}.
\end{align}
The self-energy of the electron, evaluated within SCBA, reads 
\begin{align}
\Sigma(\omega)=\frac{W^2}{3}\int \frac{d^2k}{(\frac{2\pi}{a})^2} {\left(\omega-H_{\rm SD}({\bf k}) - \delta H({\bf k}) - \Sigma+i0^{+}\right)^{-1}},
\label{eq:self}
\end{align}
where $a$ is the lattice constant. Equation (\ref{eq:self}) can be recast as  
\begin{align}
\Sigma=\Sigma_0\sigma_0+\Sigma_x\sigma_x+\Sigma_z\sigma_z+\Sigma_y\sigma_y, \label{eq:self_energy}
\end{align}
where $\Sigma_0=\frac{\Sigma_{11}+\Sigma_{22}}{2}$, $\Sigma_z=\frac{\Sigma_{11}-\Sigma_{22}}{2}$, $\Sigma_x=\frac{\Sigma_{12}+\Sigma_{21}}{2}$,
and $\Sigma_y=\frac{\Sigma_{3}-\Sigma_{21}}{2}$. 
Since the self-energy is momentum independent due to $\delta$-function disorder correlations, it modifies the parameters of the system as \cite{benakker}
\begin{align}
{\tilde m}=m+{\rm Re}\Sigma_z,~ {\tilde \omega}=\omega+{\rm Re}\Sigma_0, ~{\tilde \delta}=\delta_0-{\rm Re}\Sigma_x ,
\end{align}   
where  ``Re" refers to the real part (of the self-energy), and from now on, we focus on this part.

For typical Dirac systems, the off-diagonal part ($\Sigma_{12}$ or $\Sigma_{21}$) of the self-energy is negligibly small or zero \cite{benakker,paraj,song}. In contrast, for SD systems, $\Sigma_x$ is finite due to the anisotropic band dispersion. This can be seen by setting  $\Sigma=0$ on the right-hand side of Eq.~(\ref{eq:self}). This generates 
\begin{align}
\Sigma_{12} (\omega = 0)=-\frac{W^2a^2}{12\pi^2}\int d^2k \frac{h_x({\bf k})-i  h_y({\bf k})}{E^2({\bf{k}}) },
\label{eq:sigx}
\end{align} 
where $h_x({\bf k})$ and $h_y({\bf k})$ are defined in Eq.~(\ref{ham0}). Since $h_{x}({\bf k} )$ is an even function of ${\bf k}$, we obtain a finite
contribution to $\Sigma_x$ unlike in the isotropic Dirac scenario. Thus, disorder can  drive a topological Lifshitz  transition by renormalizing the bare $\delta_0$ for $\delta H=0$. 
In addition, for $\delta H\ne 0$, we obtain disorder-driven transitions between all three phases, the most striking being a transition from the SNT regime to the TNC regime. This signals a simultaneous transition of Lifshitz and Chern type. 

\begin{figure}
\includegraphics[width=0.49\linewidth]{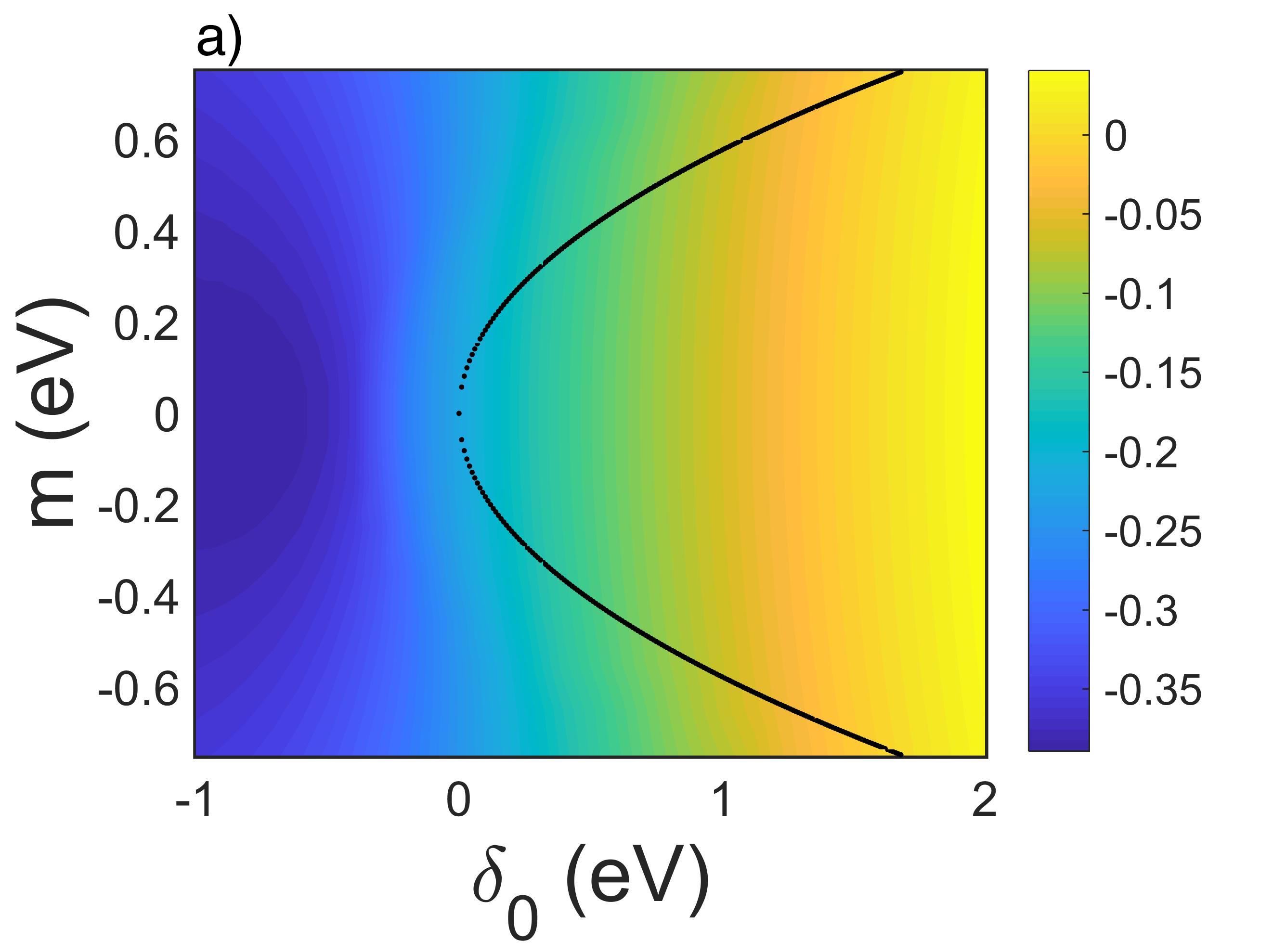}
\includegraphics[width=0.49\linewidth]{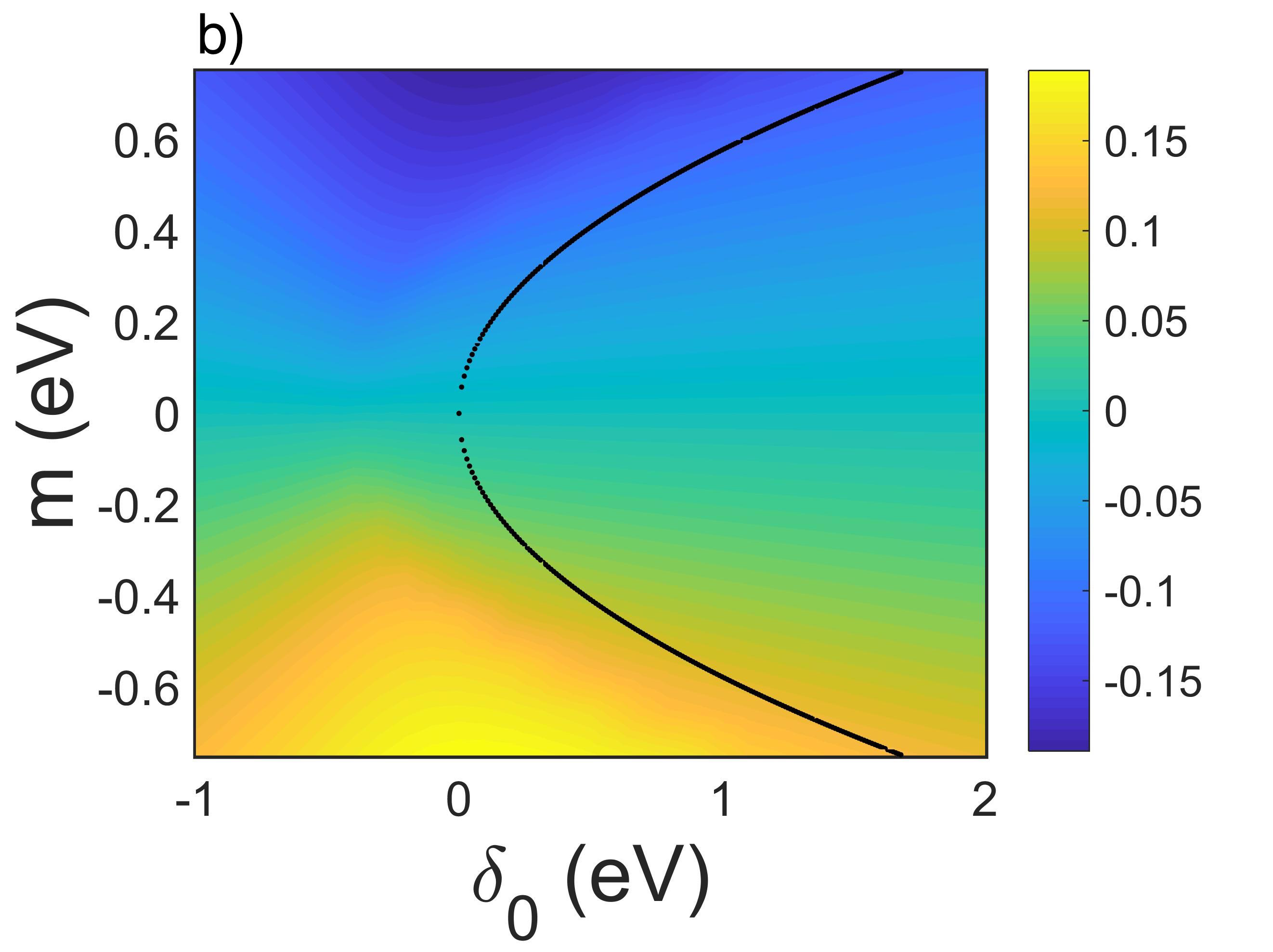}
\caption{(a) $\Sigma_x$ [Eq.~\ref{eq:self_energy}] computed for the continuum model in Eq.~(\ref{ham0}) with $\delta H\ne0$. The black dots separates the topologically trivial and non trivial phases in the clean system where $W = 0$. (b) Same as (a) for $\Sigma_z$.  Here, we have used $W = 3~eV$ with $\omega = 0$ and $\hbar^2/2\mu=0.75~ eV a^2$, $\beta=0.5~eV a$, where $a$ is the lattice constant. Note that the numerical values are chosen based on the typical parameter values of the lattice model discussed in the main text.}
\label{fig:renorm}
\end{figure}
Crucial to this transition is the relative strength and sign of $\delta_0$ and $\Sigma_x$. Figure {\ref{fig:renorm}} shows the behavior of $\Sigma_x$ and $\Sigma_z$ in the $\delta_0-m$ plane for fixed $\beta$ and disorder strength computed self-consistently using Eq.~(\ref{eq:self}). For $\delta_0<0$, $\Sigma_x$ is negative for all values  of $m$ as evident from Fig.~{\ref{fig:renorm}}. Close to the vicinity of the SNT to the TNC phase boundary, $|{\rm Re} \Sigma_x|> |\delta_0|$ leading to $\tilde \delta>0$ and thereby a transition from the SNT regime to the TNC regime occurs. However, deep in the SNT regime, $|{\rm Re} \Sigma_x|< |\delta_0|$, implying $\tilde \delta < 0$. Therefore to see a similar transition we need to crank up the disorder strength. The $\delta>0$ scenario is complicated, as $\Sigma_x$ changes sign depending on $m$ and $\delta_0$. Nevertheless, the interplay between the parameters is such that there is no transition from the TNC to the SNT regime.

Contrary to the SNT regime, the transition within the TN regime is governed by the renomalization of $m$ by $\Sigma_z$ for fixed $\beta$. The sign of $\Sigma_z$ computed self-consistently is opposite to the sign of $m$. Thus, disorder can drive states with $|m|> \beta\sqrt{2\mu\delta_0/\hbar^2}$ to states with $|m|< \beta\sqrt{2\mu\delta_0/\hbar^2}$, leading to a transition from $C=0$ to $C=1$. This follows the behavior of topological Anderson insulators \cite{li,benakker} where renormalization of mass by disorder leads to a topological phase by inverting the band.

It is worth mentioning that there exists a critical point in the $\delta_0-m$ plane, at which both $\Sigma_x$ and $\Sigma_z$ vanish, irrespective of the strength of the disorder. For $m=0$, although $\Sigma_z$ is zero for all $\delta_0$, $\Sigma_x$ vanishes at a finite $\delta = \delta_c$. This can be easily understood from Eq.~\eqref{eq:sigx}, where the contribution from $k_x^2$ coincides with that from $\delta_0$.

Finally, we point out that although the SCBA provides interesting results regarding the effect of disorder on the semi-Dirac systems, it is reliable only for weak disorder strengths. In the following sections, we study the fate of these transitions beginning with a potentially experimentally achievable microscopic lattice model, utilizing the real-space Chern number and the entanglement entropy as diagnostics for phase transitions.      

\begin{figure}
\includegraphics[width=0.42\linewidth]{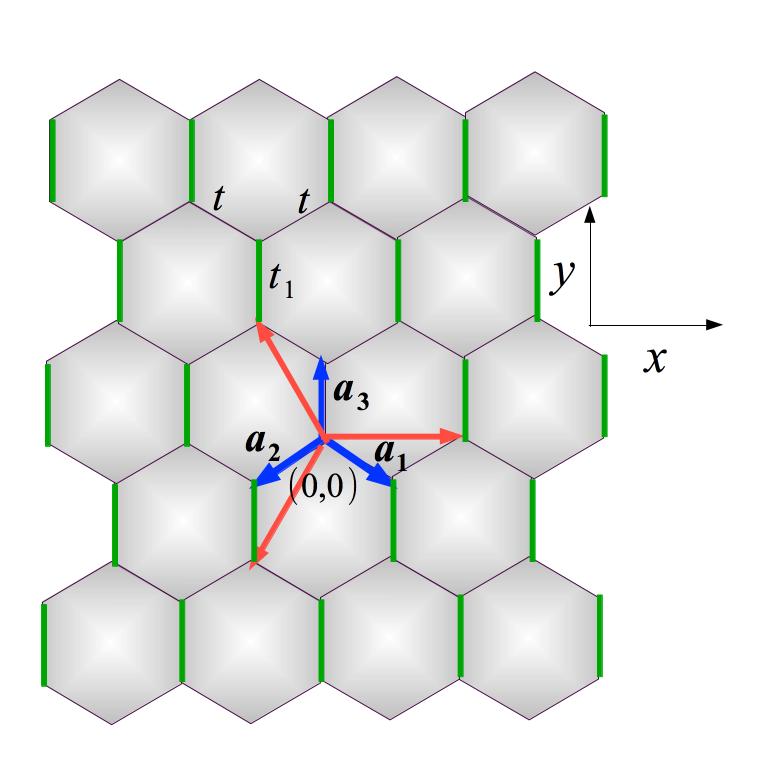}
\includegraphics[width=0.54\linewidth]{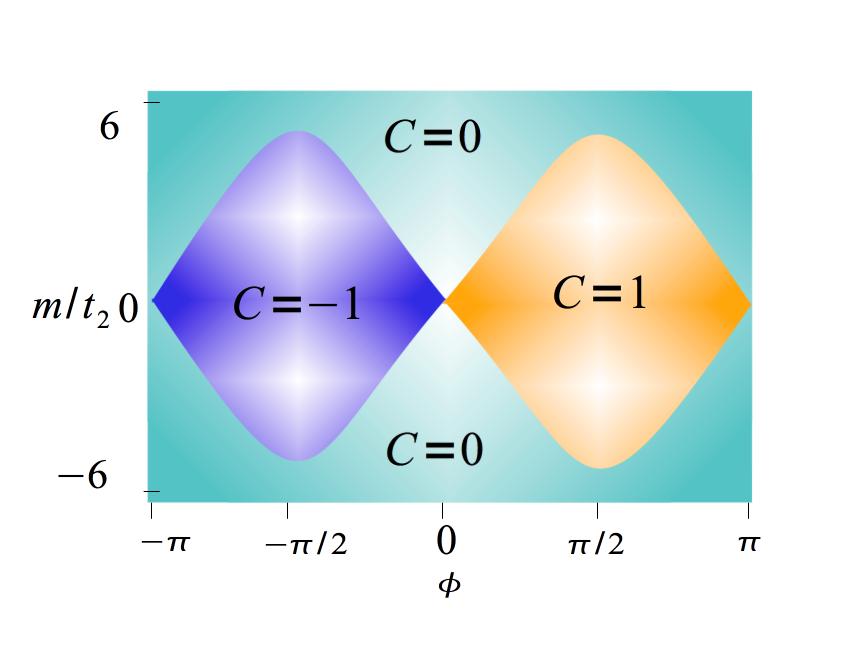}\\
\caption{(a) Lattice structure of deformed graphene. The nearest-neighbor lattice vectors (blue line) from the origin are ${\bf a_1}=\frac{a}{2}(\sqrt{3},-3), {\bf a_2}=\frac{a}{2}(-\sqrt{3},-3), {\bf a_3}=a(0,1)$, while the lattice vectors (red line) for second-nearest-neighbor atoms at angle $2\pi/3$ are given by ${\bf c_1}=\frac{a}{2}(\sqrt{3},0), {\bf c_2}=\frac{a}{2}(\sqrt{3},2), {\bf c_3}=\frac{a}{2}(-\sqrt{3},-1)$.  The positions of Dirac points in the isotropic case are given by $K(K')=\pm(\frac{4\pi}{3\sqrt{3}},0)$, while for the anisotropic case ($t_1\ne t$), they are located at $D=\left(\pm\frac{2}{\sqrt{3}}\cos^{-1}(\frac{|t_1|}{2t}),{\pi(1+sgn(t_1)) \over 3}\right)$. (b) Celebrated Haldane phase diagram as a function of flux $\phi$ for $t_1=t$. For the results in the main text, we fix $\phi=\frac{\pi}{2}$ and investigate the effect of disorder on the topological phases of the deformed graphene.}
\label{fig:graphene}
\end{figure} 

\section{Phases in an anisotropic Honeycomb Model}
We now study a hexagonal lattice model with anisotropic hopping which is well known to host a semi-Dirac dispersion and a Lifshitz transition \cite{montam}. Consider the tight-binding model of graphene with nearest- and next-nearest-neighbor hopping \cite{haldane} ,
\begin{align}
\mathcal{H} = \sum_{\langle ij\rangle}t_{ij}c_{i}^{\dagger}c_j+\sum_{\langle\langle ij\rangle\rangle}t_2 e^{i\phi}c_{i}^{\dagger}c_{j} 
+m \sum_{i} (-1)^i c_{i}^{\dagger}c_{i},
\label{eq:tight}
\end{align}
where $c_i (c_i^{\dagger})$ annihilates (creates) an electron at site $i$, $t_{ij}$ is the nearest-neighbor hopping, $t_2$ is next-nearest-neighbor hopping, $m$ is the onsite staggered potential and $\phi$ is the phase acquired by $t_2$ due to a periodic magnetic field. Considering a different hopping along the $y$-direction, $t_1$, compared to the other directions, $t$ Fig.~\ref{fig:graphene}, Eq.~(\ref{eq:tight}) can be recast in momentum space as   
\begin{align}
\mathcal {H}({\bf k})=\tilde h_x({\bf k} )\sigma_x+\tilde h_y({\bf k})\sigma_y+M({\bf k})\sigma_z,
\end{align}
 \begin{figure}
\includegraphics[width=0.49\linewidth]{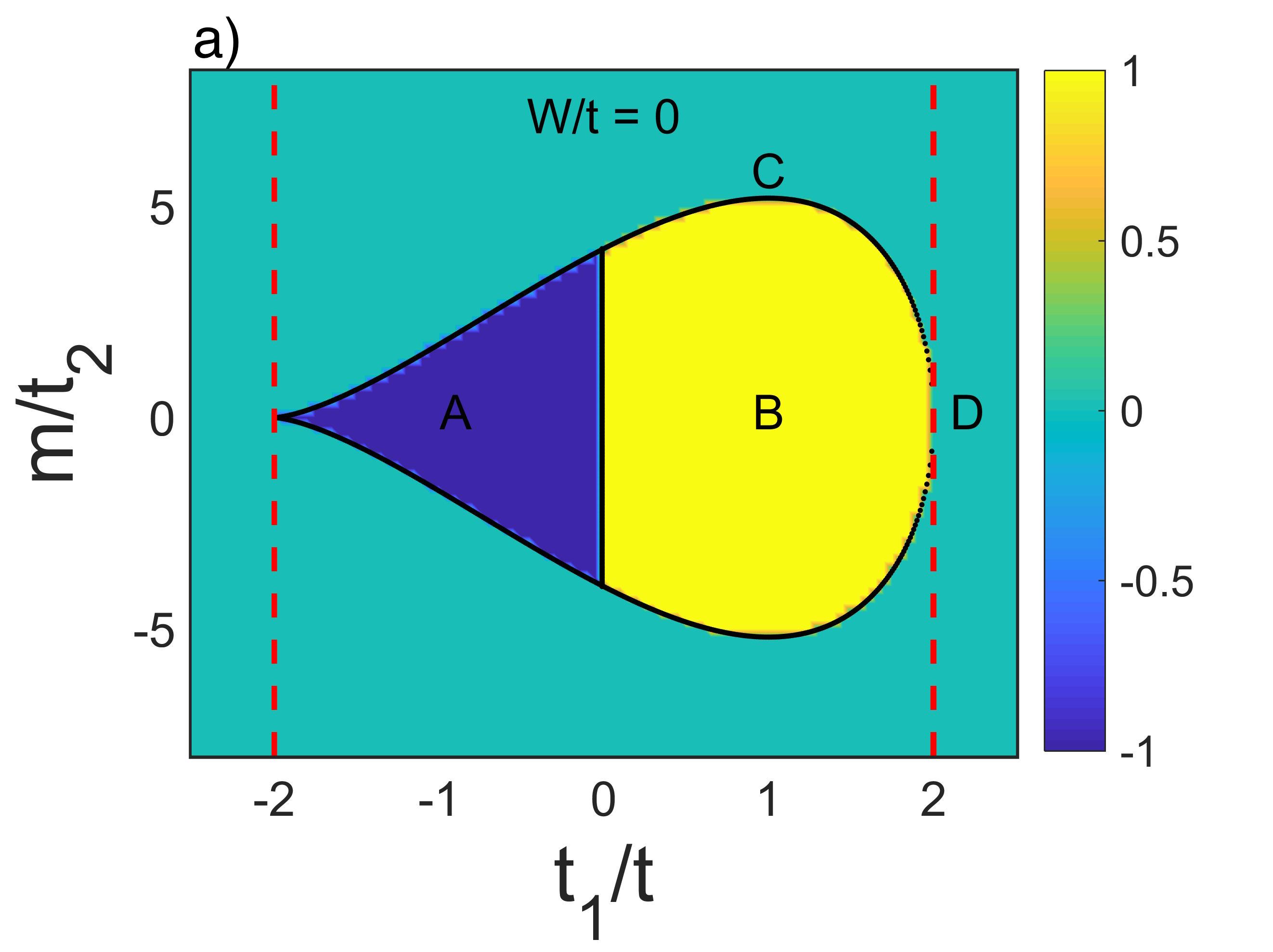}
\includegraphics[width=0.49\linewidth]{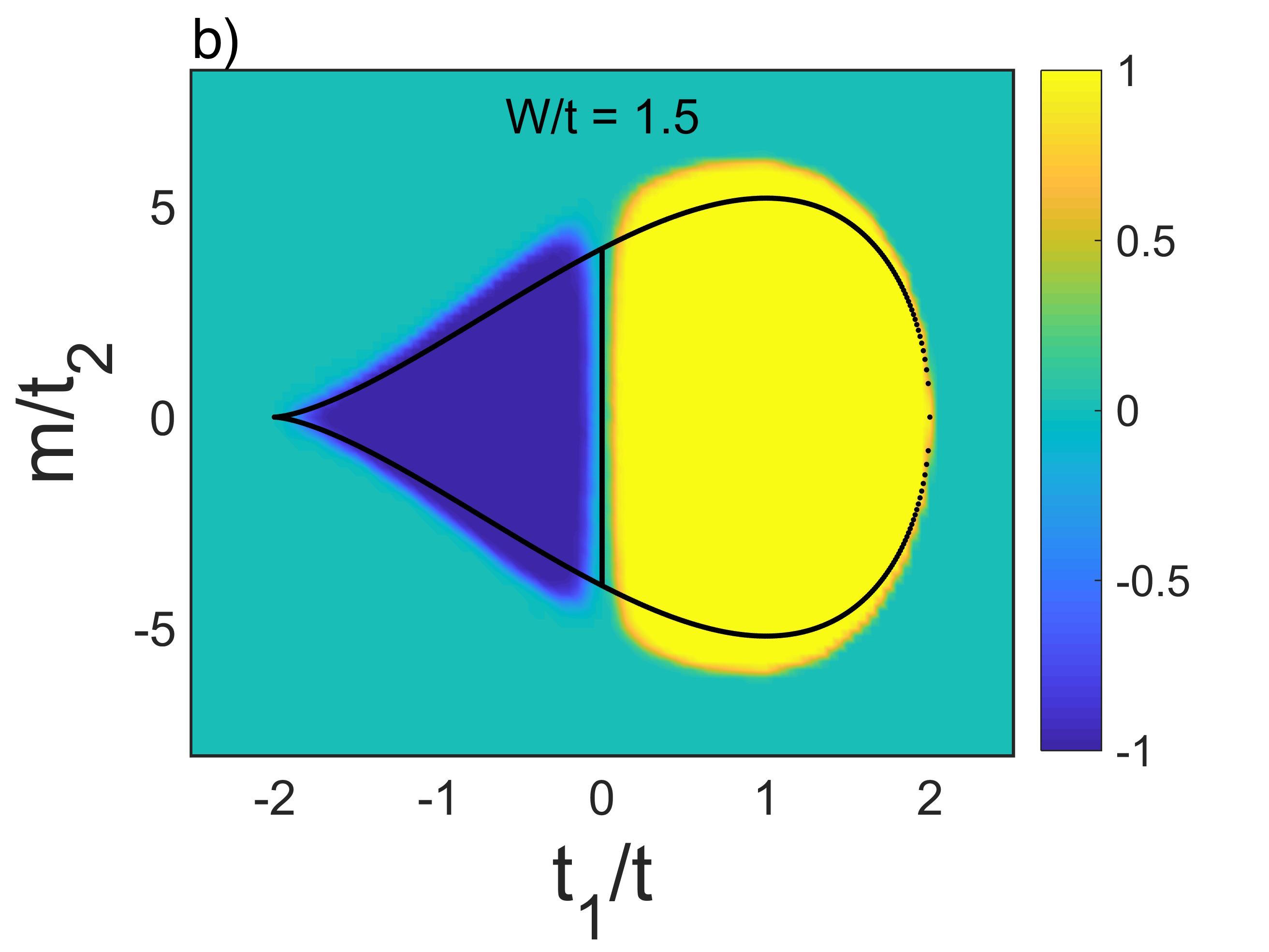}\\
\includegraphics[width=0.49\linewidth]{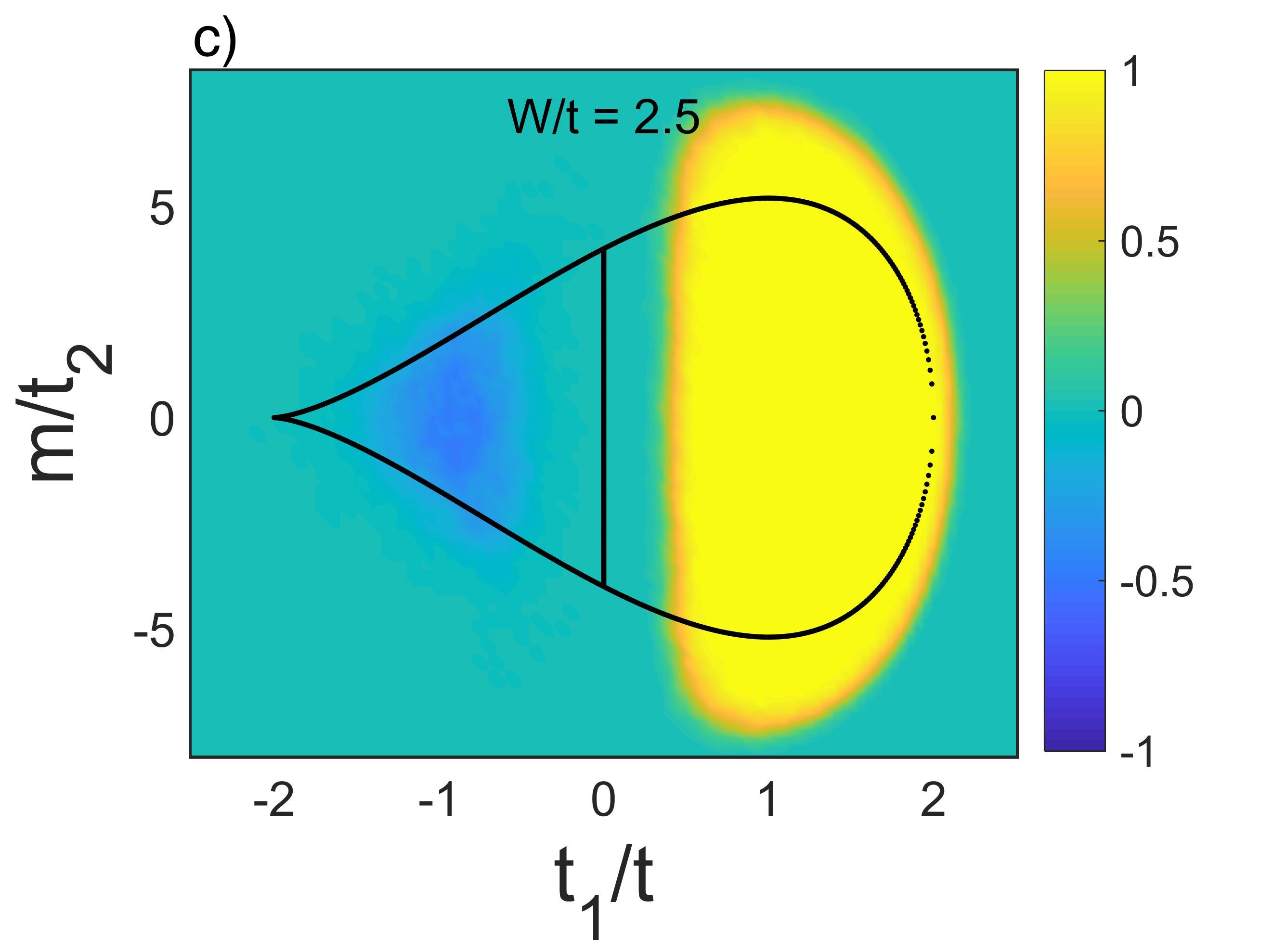}
\includegraphics[width=0.49\linewidth]{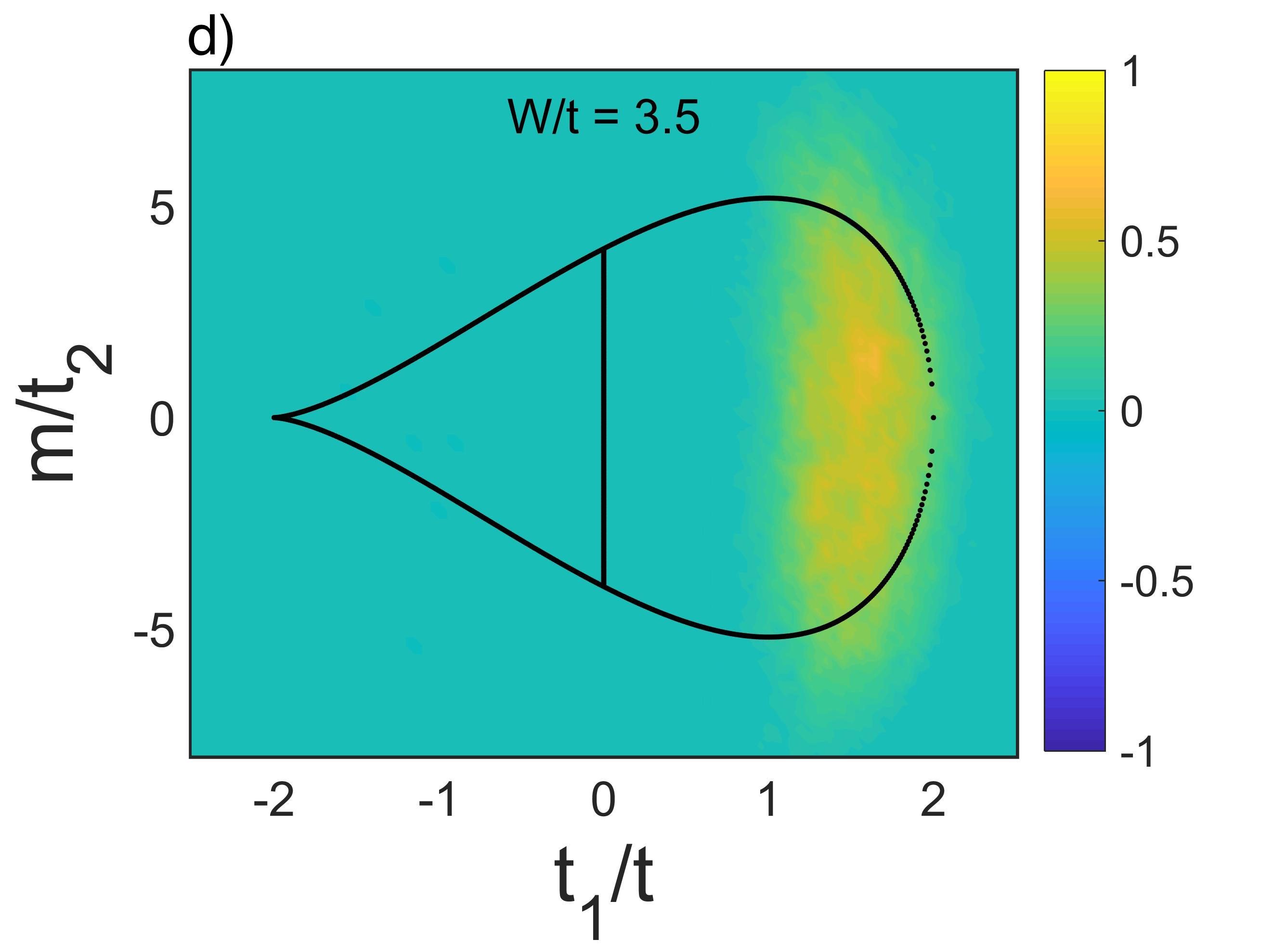}
\caption{Evolution of the phase boundary as a function of disorder strength for $\phi = \pi/2$, $t_2/t = 0.25$. (a) In the clean limit, the numerically calculated phase boundary matches the phase boundary (black dots) found analytically.  The regime within the red-dashed line represents the TN while that outside denotes the SN regime. The four points labeled A-D in this diagram are discussed in the main text. (b--d) Phase diagram for different disorder strengths. For weak disorder the phase boundary within the two-node regime enhances as a signature of topological states induced by disorder. However, for strong disorder, the topological states are destroyed by localization.}
\label{fig:disorder}
\end{figure}
where 
\begin{align}
&\tilde h_x({\bf k})=t_1+2t\cos\left(\frac{\sqrt{3}}{2} k_x a\right)\cos\left(\frac{3}{2} k_y a\right)\nonumber\\
&\tilde h_y({\bf k})=2t\cos\left(\frac{\sqrt{3}}{2} k_x a\right)\sin\left(\frac{3}{2} k_y a\right)\nonumber\\
&M({\bf k})=m - 2t_2 \sin\phi ~ \left[ \sin\left(\frac{\sqrt{3}}{2} k_x a + \frac{3}{2} k_y a\right) \right.\nonumber\\
&~~~~~~+\left. \sin\left(-\frac{\sqrt{3}}{2} k_x  a-\frac{3}{2} k_y a\right)+\sin\left(\sqrt{3} k_x a \right)\right].
\label{eq:gra}
\end{align}
\begin{figure}
\includegraphics[width=0.99\linewidth]{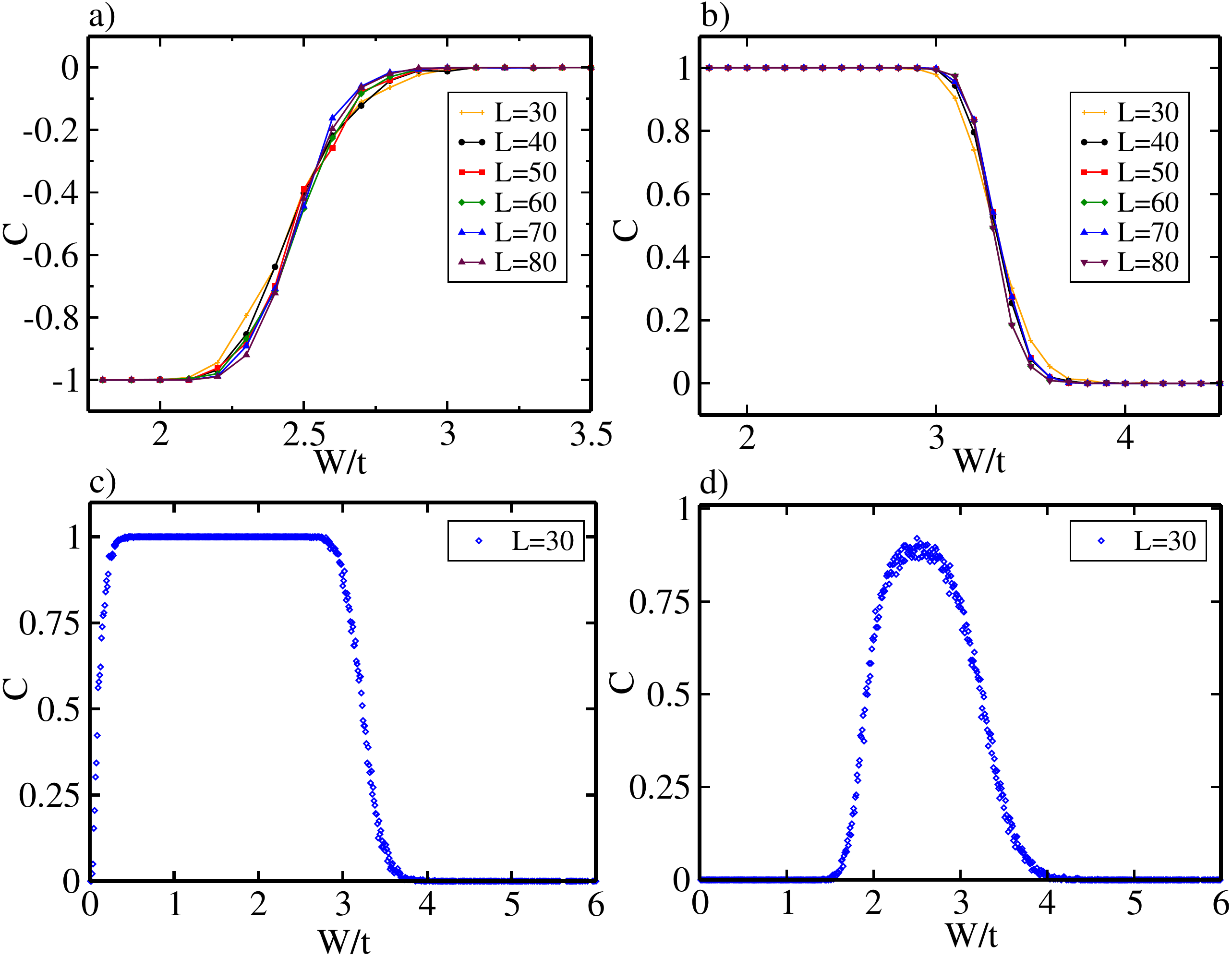}
\caption{Variation of Chern number as a function of disorder strength, $W$, starting from a trivial or topological phases as indicated in Fig.~\ref{fig:disorder}(a). (a, b) Effect of system size on the disorder-averaged Chern number with disorder strength in the TNC regime starting from points {\bf A} (panel a: $t_1/t = -1, m/t_2=0.2$ ) and {\bf B} (panel b: $t_1/t = 1,m/t_2 =0.2$). (c) Starting from point {\bf C} (panel c: $t_1/t = 1, m/t_2=5.2$), disorder induces topological phases from an insulating phase with $C=0$ (cf. Fig.~\eqref{fig:clean}). (d) Same plot as (c) for point {\bf D} (panel d: $t_1/t = 2.05, m/t_2 =0.2$) in the single-node trivial regime. From the SNT, disorder drives the transition to the TNC regime with Chern number $C=1$.}
\label{fig:clean}
\end{figure}
For $M(k)=0$, the locations of Dirac points are given by $D=\left(\pm\frac{2}{\sqrt{3}}\cos^{-1}(\frac{|t_1|}{2t}),\frac{\pi (1+sgn(t_1))}{3}\right)$. At $|t_1|=2t$, the two Dirac points merge at $(0,\frac{\pi (1+sgn(t_1))}{3})$. Expanding $\tilde h_x({\bf k})$ and $\tilde h_y({\bf k} )$ around this point, we obtain 
\begin{align}
&\tilde h_x({\bf k})=-(2t-|t_1|)+\frac{3t a^2}{4}k_x^2\nonumber\\
&\tilde h_y({\bf k})=3t ak_y. 
\label{eq:latcontm}
\end{align}
Note that, $|t_1|>2t$ corresponds  to a gapped insulating phase, while $|t_1|<2t$ denotes semi-metallic phase with two Dirac nodes. Also, comparing ${\bf h(k)}$ in Eq.~(\ref{ham0}) with $\tilde h {(\bf k)}$ of Eq.~(\ref{eq:latcontm}), we obtain $\delta_0= 2t-|t_1| $, $\frac{\hbar^2}{2\mu}= \frac{3t a^2}{4}$ and $\hbar v_F= 3t a$. 

In Eq.~(\ref{eq:gra}), $t_1=t$ corresponds to the celebrated Haldane model with distinct topological phases as a function of $\phi$ as shown in Fig.~(\ref{fig:graphene})b.  However, for fixed $\phi$, we find a rich topological phase diagram as a function of $t_1$. The corresponding plot is shown in Fig. ~(\ref{fig:disorder})a with $\phi = \frac{\pi}{2}$. The phase diagram can be divided into two distinct regimes as in the continuum model.  The regime $|t_1| \le2t $ can be identified as the TN regime, while $|t_1|>2t$ can be identified as the SN regime. Thus the transition from TN to SN corresponds to the Lifshitz transition in conjunction with the continuum limit. In the TN regime, three distinct topological phases appear with topological invariants $C=0,\pm1$ (Fig.~\ref{fig:disorder}a). The phase boundary separating $C=0$ from $C=\pm 1$ is given by
\begin{align}
\frac{m}{t_2}=\pm2\sin\phi\left[ -2 \sqrt{1-\frac{t_1^2}{4t^2}} + \sin\left(2\cos^{-1}\left(-\frac{t_1}{2t}\right)\right) \right].
\label{eq:phas}
\end{align}
On the other hand, the two topological phases, $C=\pm 1$, are separated by a line segment determined by $t_1=0$ and $-4\le m/t_2\le 4$. Along this line the band spectrum shows two gapless points at $k_x=\pi/\sqrt{3}$ and $k_y= \pm\frac{2}{3} \cos^{-1}(-\frac{m}{4 t_2})$. Moreover, they are distinct topological phases not related by time-reversal symmetry unlike in the Haldane model. The phase boundaries for these two distinct topological states are asymmetric with respect to $t_1=0$ due to the anisotropic band dispersion.  Note that, the Chern phase boundary is maximum for $\phi = \pm \pi/2$ and changing $\phi$ squeezes the boundary along the $m/t_2$ direction. 

\begin{figure}
\includegraphics[width=0.99\linewidth]{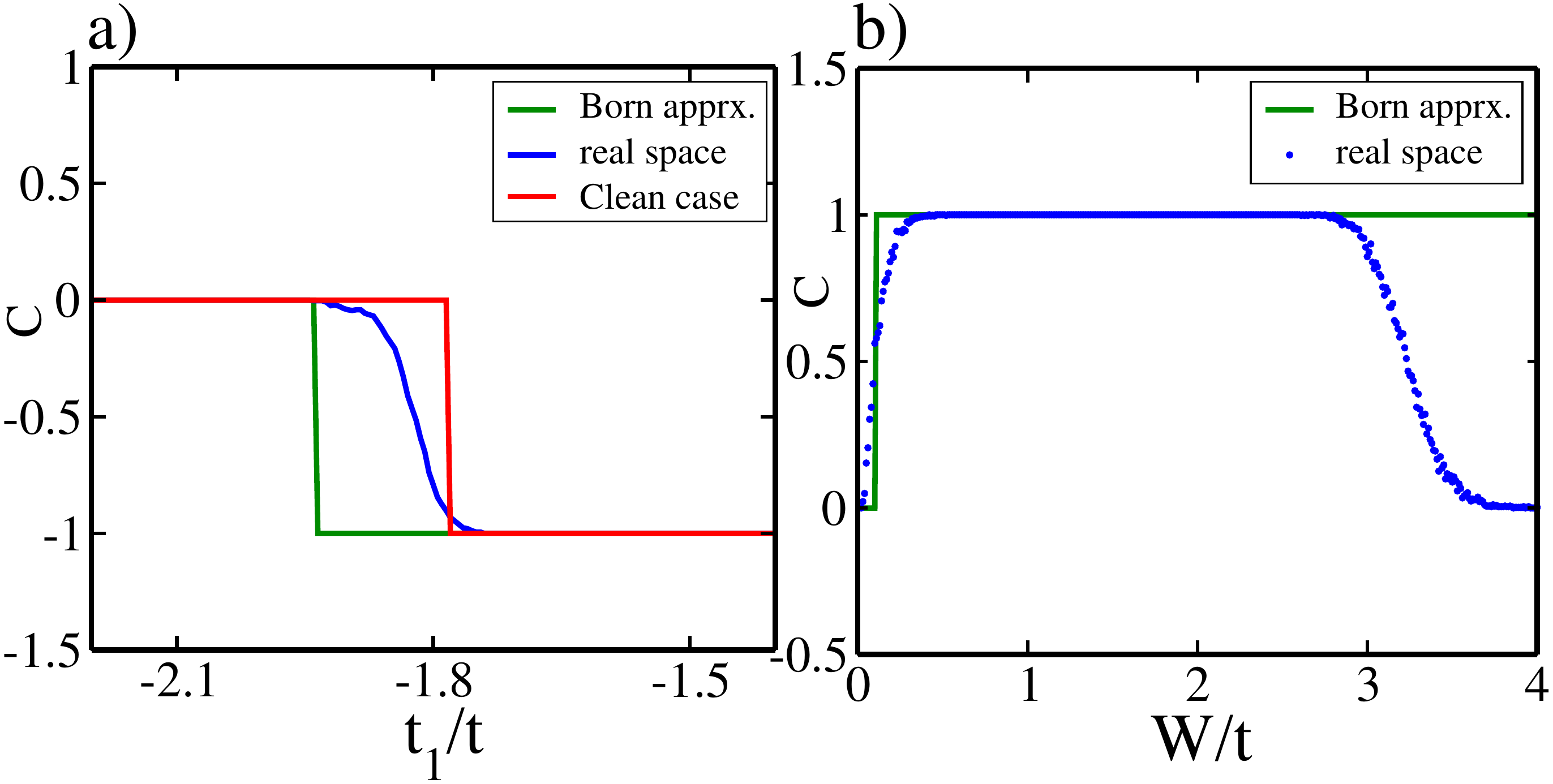}
\caption{Comparison between Chern number calculated from the renormalized parameters (using SCBA) and Chern number from the real space calculation.
(a) Chern number as a function of $t_1/t$ for fixed $m/t_2=0.2$ and $W/t=1.0$.  Clearly, there is a qualitative agreement between the two methods mentioned above. 
For comparison, we also plot the Chern number in the clean limit. (b) Variation of Chern number as a function of disorder strength, $W/t$ for $t_1/t = 1, m/t_2=5.2$. For strong disorder, there is a significant disagreement between the two methods in conjunction with the discussion in the main text.  Note that here we have used a lattice model for calculating the Born approximated Chern number.}
\label{fig:compare}
\end{figure}

\section{Disorder-averaged real space Chern number}
To address the effect of disorder, we add an onsite disorder potential to  Eq.~(\ref{eq:tight}) 
\begin{align}
\mathcal{H_{\rm dis}} & =\sum_i \mu_i c_i^\dagger c_i,
\end{align}
where $\mu_i$ is picked randomly from a uniform distribution $\left[-W,W\right]$.
Following  Ref.~[ \onlinecite{sheng}], we compute the real-space Chern number in 
the presence and absence of disorder.
In the clean limit, $W = 0$, the numerically calculated phase boundary, with total number of sites $2L\times L = 2\times 30\times 30$, for different topological phases [cf. Fig.~\ref{fig:disorder} (a)] matches well with the phase boundary predicted from the analytical results in Eq.~(\ref{eq:phas}) and the line of gapless points. 
With disorder, the phase boundaries computed for $150$ disorder configurations shifts, reflecting
disorder-driven topological transitions [Figs.~\ref{fig:disorder}(b)-\ref{fig:disorder}d] .
\begin{figure}
\includegraphics[width=0.99\linewidth]{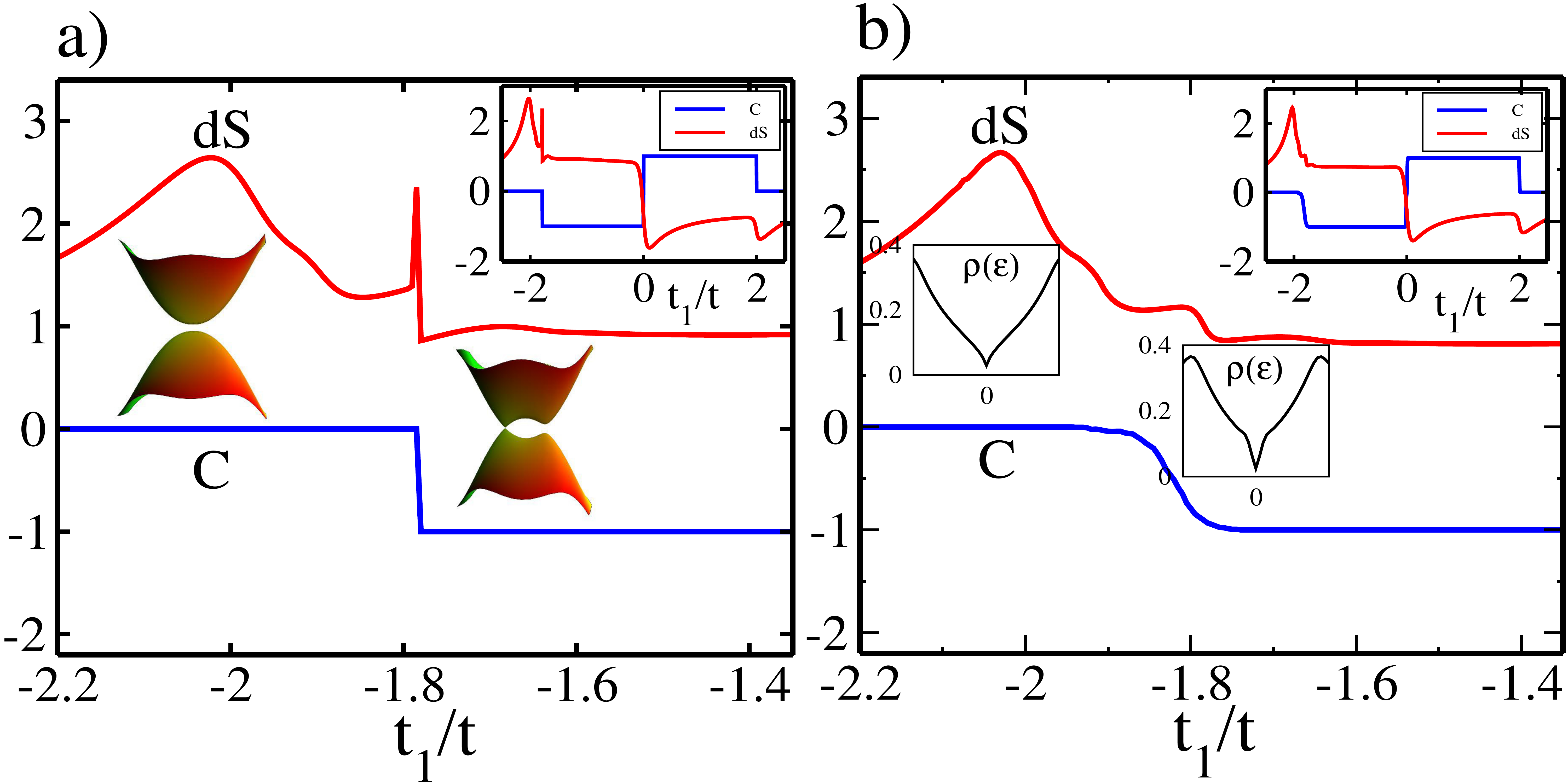}
\caption{(a) Derivative of entanglement entropy ($dS$ in arbitrary units) as a function of $t_1/t$ for fixed $m/t_2=0.2$ in the clean limit ($W=0$) with system size $L = 60$. 
The peak at $t_1/t\sim-1.78$ corresponds to the change in the Chern number, while the peak at $t_1/t\sim -2$ is associated with the change in 
band curvature as indicated in the plot [Inset: zoomed-out view for the range of $t_1/t$ as in Fig.~\ref{fig:disorder}(a)].
 (b) Same as (a) for disorder strength $W/t=1$ and system size $L = 40$. 
Note that the peak structure is unchanged even for finite disorder. However, the peak associated with the Chern number
shifts, reflecting the shifts in the Chern phase boundary as discussed in the main text. 
(For completeness, disorder-averaged density of states are shown near the topological and Lifshitz transition points.)}
\label{fig:ee}
\end{figure}

Fig.~\ref{fig:disorder}(b) and \ref{fig:disorder}(c) show the phase boundary in the presence of weak disorder. Clearly, the shift in the phase boundaries results in the appearance of topological states  both in the single- and two-node trivial insulating regimes. Similar to the continuum model, the most interesting result here turns out to be disorder-driven topological states in the SNT regime which is intrinsically trivial even in the absence of time-reversal symmetry. The emergence of such states can be attributed to the renormalization of hopping, which in turn renormalises $2t-t_1$. In contrast, the topological states in the TNT regime appears due to the renormalization of $m/t_2$. Thus, the results obtained from the lattice model corroborate the results of the self-consistent Born-approximation in the continuum for weak disorder strengths. For stronger disorder, the phase boundaries diminish [Fig.~\ref{fig:disorder}(d)], as opposed to SCBA results, which predict the SNT to TN transition even for sufficiently strong disorder. For concreteness, in Fig.~(\ref{fig:compare}), we compare the disorder-averaged real space Chern number with the Chern number obtained from the renormalized  parameters using the SCBA. Evidently, there is qualitative agreement between these two methods in the weak disorder limit.

To elaborate the above transitions, in Fig.~\ref{fig:clean}, we show variation of Chern number as a function of disorder strength, starting from four different points ({\bf A, B, C, D}) in the phase space as indicated in Fig.~\ref{fig:disorder}(a). {\bf A, B, C} belong to the TN regime with three distinct topological phases $C=-1,+1,0$, respectively, while {\bf D} belongs to the SN regime with $C=0$. For the case {\bf A}, the non-zero Chern phase survives until $W/t=3.1$, while for {\bf B}, it survives until $W/t = 3.8$ [Fig.~\ref{fig:clean}(a)-\ref{fig:clean}(b)]. Notice that, there is no significant variation in the Chern number if we vary the size from $L = 30$ to $L = 80$ in a system size of $2L\times L$. Thus, for {\bf C} and {\bf D}, we present results only for   $L=30$. In contrast to ({\bf A  B}),   ({\bf C} and {\bf D}) correspond to a trivial insulating phase in the clean limit. Disorder introduces a non-zero Chern number, thereby leading to a development of a Chern phase. Fig.~\ref{fig:clean}(c) and \ref{fig:clean}(d) evidence the appearances of such phases. Despite the proximity of the points to the topological phases, the appearance of Chern phase in {\bf D} requires stronger disorder than {\bf C} as expected from the earlier discussions. Likewise, the topological states in {\bf D} disappear before {\bf C} with increasing disorder. 

Next, we comment on the non-integer values of Chern numbers appearing in Fig.~\ref{fig:clean}. A non-integer Chern number in a disordered system is an effect of finite system size and a finite averaging of disorder configurations. With increasing system size and number of disorder configurations, the Chern numbers start to move towards the quantized value. However, this comes at huge computational cost of the method in practice used to calculate the size dependent Chern number. Thus, we find non-integer Chern numbers for all the cases near the topological transition (Fig. 6). However, in the case of D, the Chern number never reaches the quantized value as a function of disorder strength. This may be attributed to the robustness of the insulating phase for $t_1/t>2$. We would indeed require large system size and number of disorder configurations to resolve the Chern states in this regime.

\begin{figure}
\includegraphics[width=0.9\linewidth]{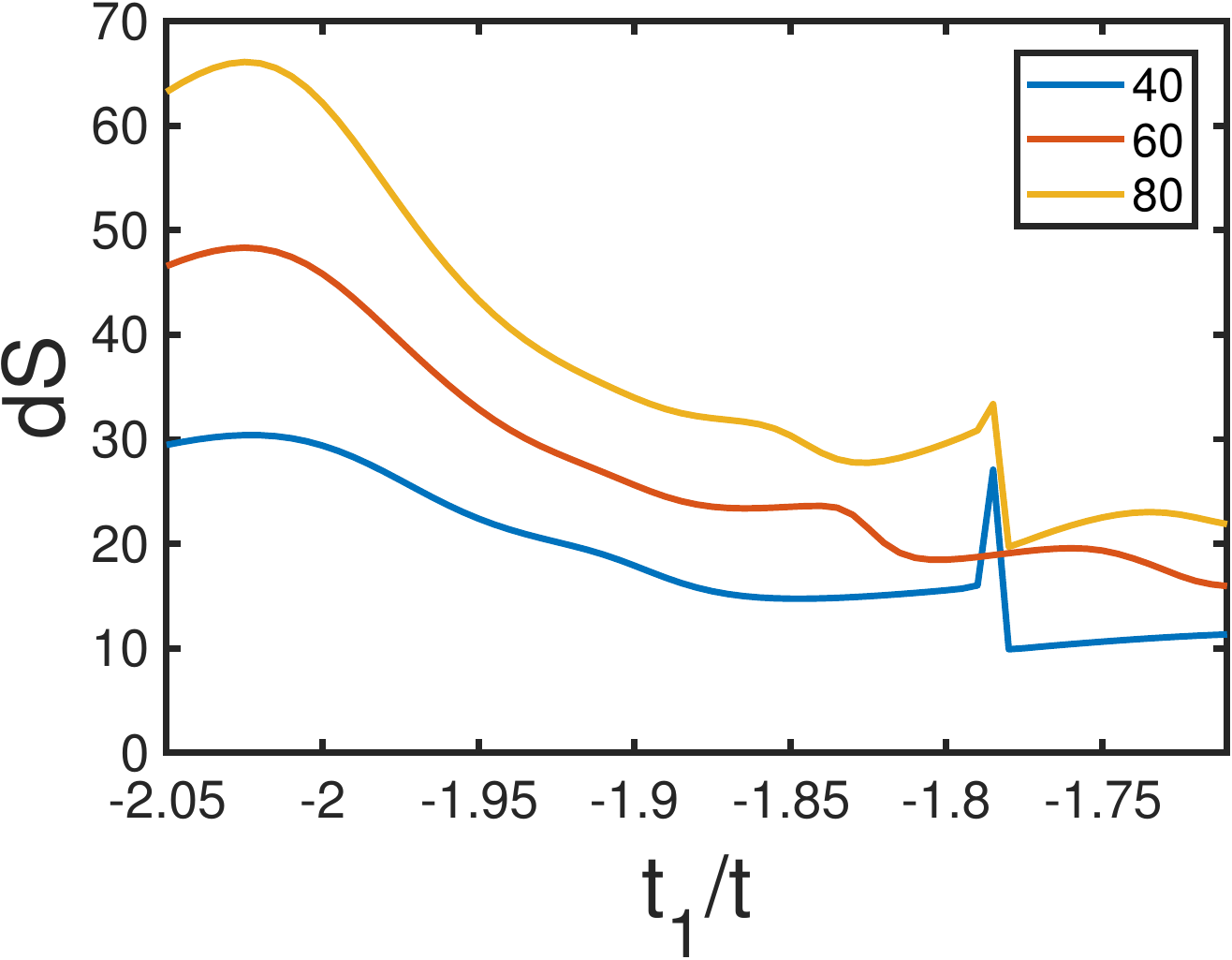}
\caption{Derivative of entanglement entropy in the clean limit for different system size. 
The peak near $t_1/t\sim -1.78$ corresponds to the topological transition, whereas the peak near $t_1/t\sim -2 $ corresponds to a change in 
the band curvature.}
\label{fig:entropyL}
\end{figure}

\section{Disorder-averaged entanglement entropy and Density of states}
It is well known that entanglement entropy (S) in disordered systems is a tool to capture topological transitions. Specifically, it has been shown that the derivative of disorder-averaged entanglement entropy \cite{tami,tami1,huges1,huges2,vojta,pour} with respect to the parameters shows peaks near the transition as a consequence of gap closing. 
We therefore verify topological transitions discussed in the preceding section by calculating S, which is defined as $S =  -Tr(\rho_B \ln \rho_B)$, where $\rho_B$ is the reduced density matrix of half of the sub-system B of original system. 

Figure~\ref{fig:ee}(a) shows the numerical derivative of the entanglement entropy, $dS = \frac{\Delta S}{\Delta t_1}$, as a function of $t_1$, in the clean limit. For comparison, we also show variation of Chern number with $t_1$. We obtain multiple peaks in the derivative of $S$. For brevity and clarity, we focus on the $t_1/t<0$ regime. The peak near $t_1/t\sim -1.78$ evidences a topological transition from $C=0$ to $C=-1$ as a manifestation of the gapless point at the transition. However, the sharpness of this peak depends on the system size due to a commensurability effect between the momentum values set by the finite system size and the momentum at which the gap closes. This is apparent in Fig.~\ref{fig:entropyL}, where we find relatively sharp peaks only for $L=40$ and $L=80$. 

The peak near  $t_1/t\sim-2$ can be traced back to the change of the band curvature (merging of Dirac points) or Lifshitz transition. For $m=0$, the merging of two Dirac points (TN to SN Lifshitz transition) and the Chern transition coincide, which in turn leads to a single peak in $dS$ at $t_1/t=-2$. For finite $m$, the Chern transition shifts from $t_1/t\sim-2$, as evident from the phase diagram in Fig.~\ref{fig:disorder}(a). Consequently, the single peak in $dS$ dissociates. Thus for small but finite $m$, the peak at $t_1/t\sim-2$ corresponds to the Lifshitz transition, and it consistently grows with system size within the scope of our numerical calculations (Fig.~\ref{fig:entropyL}). However, it flattens as $m$ increases. This is because all band features become less sharply defined as the gap increases. Note that for finite sizes, the appearance of a peak due to the Lifshitz transition is consistent with a recent study \cite{sun-sik} which discusses entanglement entropy as a probe to detect Lifshitz transition/Fermi surface topology for a gapless system. 
 
\begin{figure}
\subfigure[]{
\includegraphics[width=0.45\linewidth]{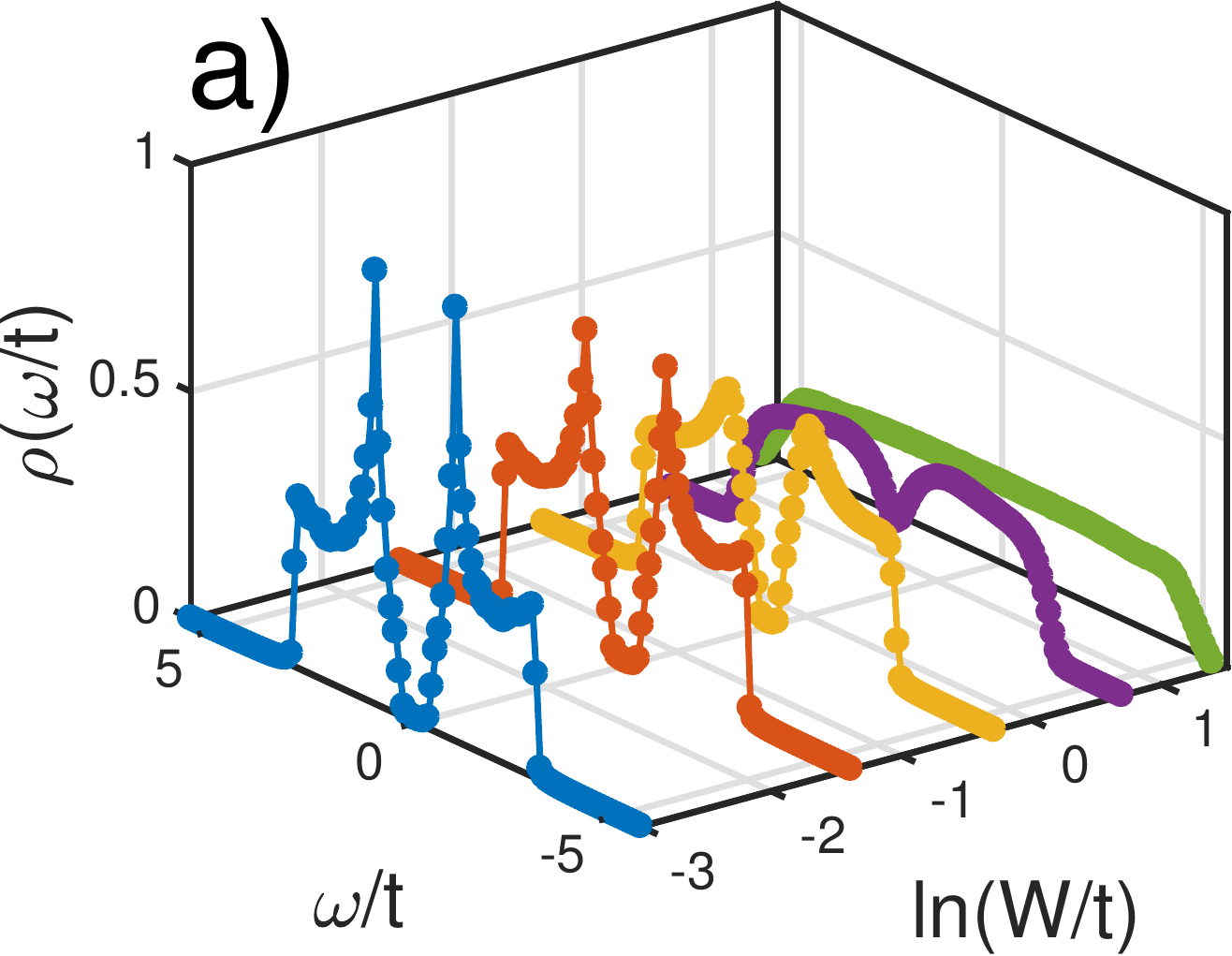}}
\subfigure[]{
\includegraphics[width=0.45\linewidth]{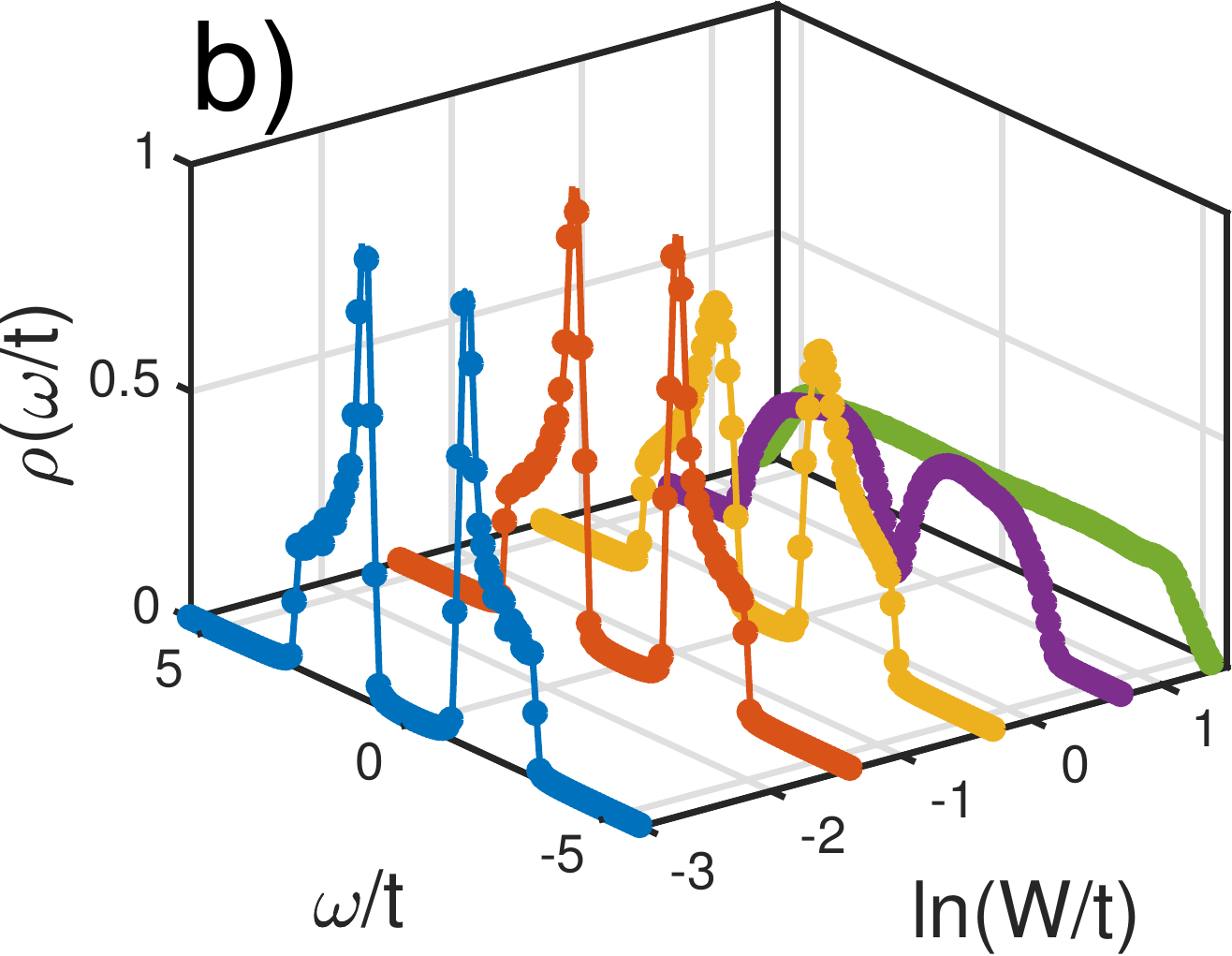}}
\subfigure[]{
\includegraphics[width=0.45\linewidth]{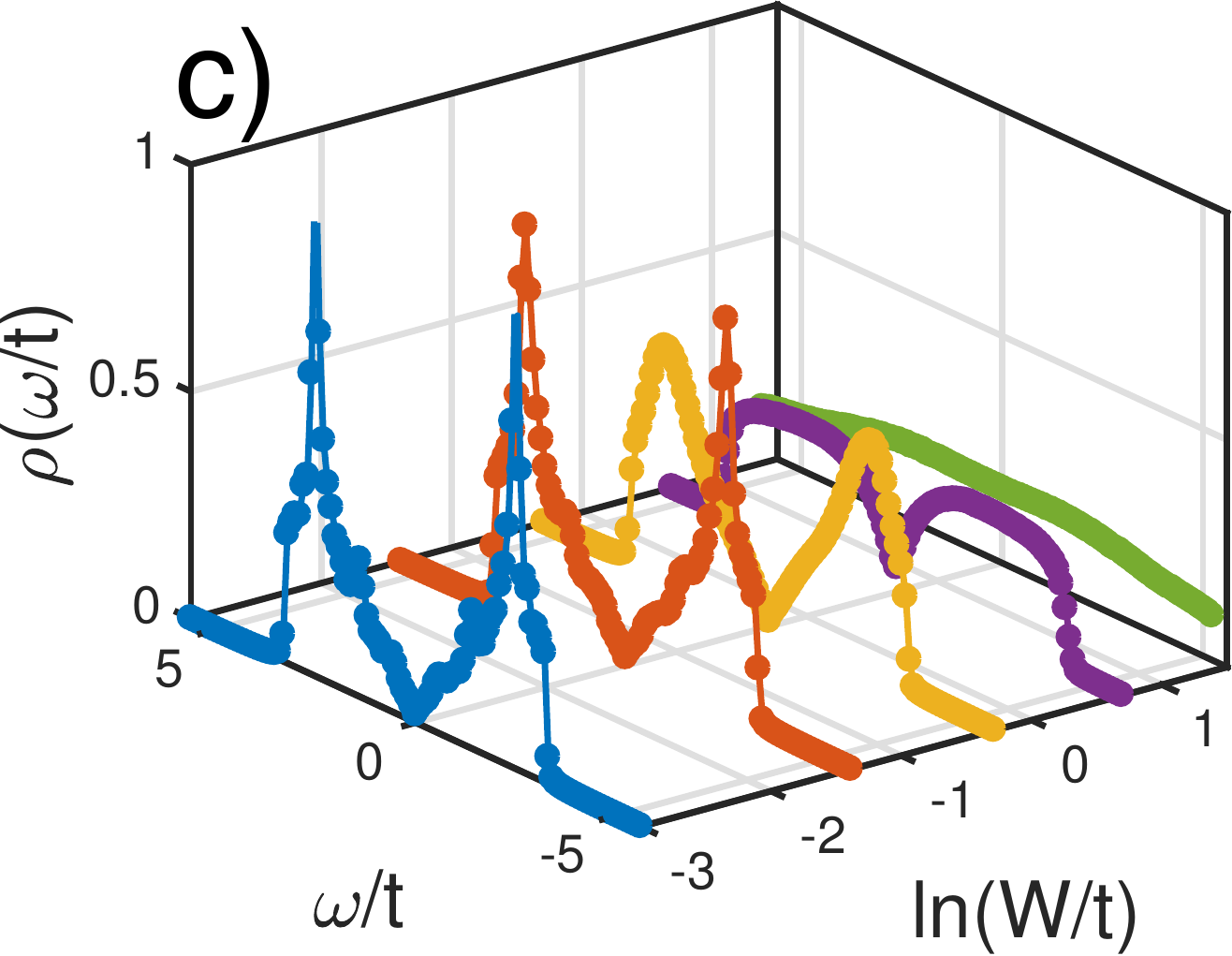}}
\subfigure[]{
\includegraphics[width=0.45\linewidth]{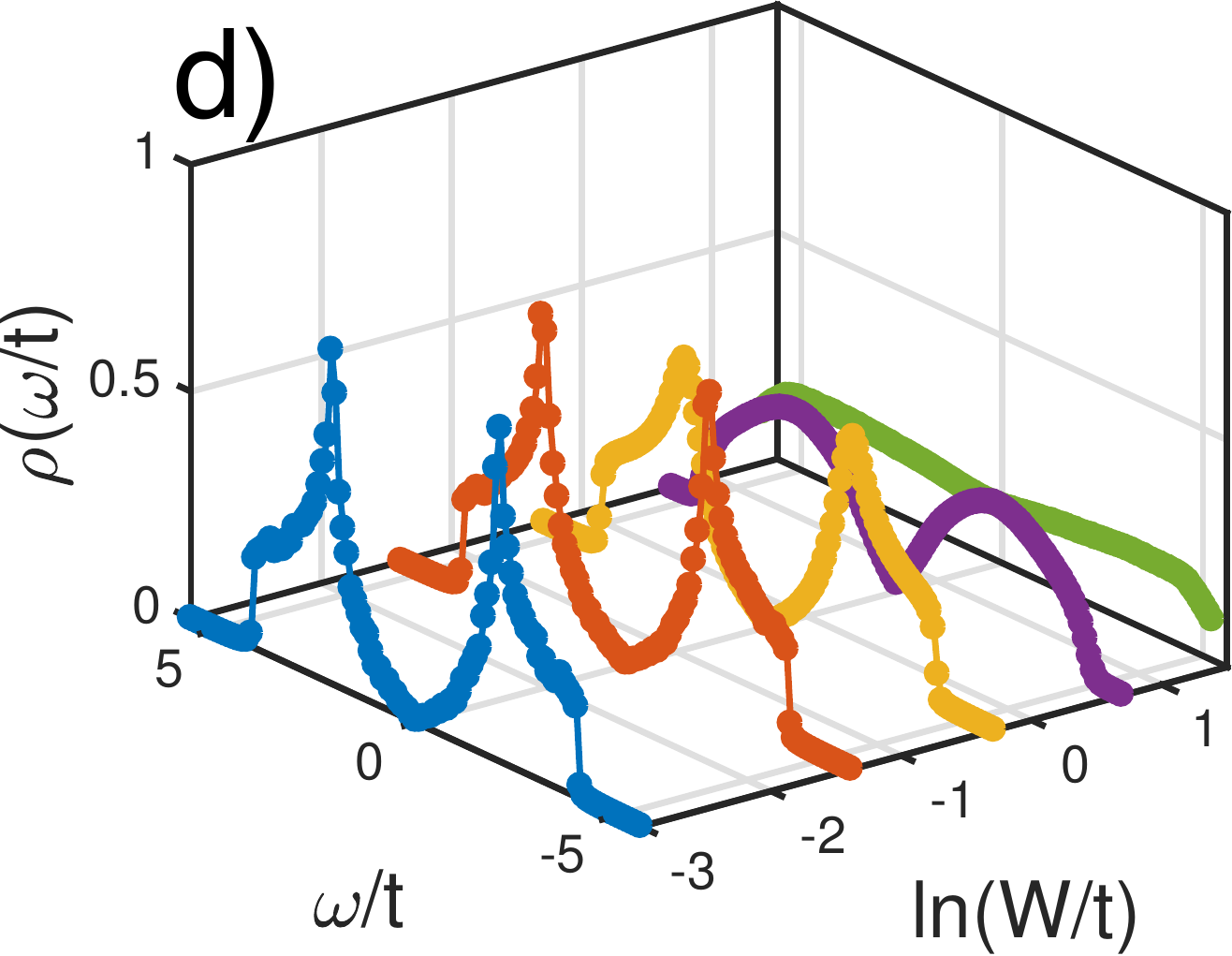}}
\caption{The density of states for parameters : $\phi = \pi/2$, $t_2/t = 0.25$ for different disorder strengths $W/t = 0,0.25,0.75,2,4$. The top panel corresponds to points {\bf A}, {\bf B} in Fig.~\ref{fig:clean}, whereas the bottom panel corresponds to {\bf C} and {\bf D} in the same figure. The clean limit curve has been shifted for visual purposes. The number of disorder configurations is $500$ with system size $L = 60$. }
\label{fig:dos_lattice}
\end{figure}

Fig.~\ref{fig:ee}(b) displays the disorder-averaged derivative of the entanglement entropy for $W/t=1$, computed for $500$ disorder configurations and $L=40$. The peaks survive in the presence of weak disorder. However, they start to disappear with stronger disorder due to localization. 

Before ending this section, we briefly discuss the disorder-averaged DOS, calculated from the lattice model for different disorder strengths. 
Figure~\ref{fig:dos_lattice} illustrates the DOS for the four different points of interests ({\bf A}, {\bf B}, {\bf C}, and {\bf D} ) in the phase diagram of Fig.~\ref{fig:disorder} (a).
In the clean limit, the DOS of ({\bf A}, {\bf B}, and {\bf C}) show the saddle-point nature of the two-node dispersion close to the Fermi energy which is absent in {\bf D} which belongs to the single-node regime. But the DOS does not show any signature that can help locate the transition associated with the Chern number, as is expected. Disorder initially closes the gap between the bands and smears out the Van Hove singularities, thereby erasing any of the remaining band features. For large disorder the extended states contributing to the Chern number annihilate and the system becomes localized \cite{castro2016,masaru2007}. At this stage we expect no significant difference in the density of states of the four points as seen in Fig.~\eqref{fig:dos_lattice}. Thus, entanglement entropy can be considered as a useful diagnostic for both topological and Lifshitz transitions rather than the DOS of the system.

\section{Experimental Realization}
In this section, we discuss the scope of experimental realization of the physics discussed in the preceding sections. Anisotropic graphene with $t_1 \ne t$, namely, quinoid, has been discussed long ago by Pauling \cite{paul}, and later it was shown that such an anisotropic situation can be induced by uniaxial stress or bending of a graphene sheet \cite{ryad}. However, controllable variation of $t_1$ to access various regimes of interest may not be easily achievable in deformed graphene or any other proposed organic materials that host such dispersion. Thus the most promising platform to look for such lattice structure is cold atoms trapped in a laser-induced hexagonal optical lattice. It has been recently shown in Ref.~[\onlinecite{tilman}] that Dirac points in a hexagonal optical lattice can be moved and merged to have a semi-Dirac dispersion. Thus in this optical setting, variation of $t_1$ can be  achieved by applying an oscillatory gauge field $A_0\sin(\omega t)$ along the $y$-direction as shown in Fig.~\ref{fig:graphene}. This gauge field modifies $t_1$ as $t_1 \int dt ~e^{i A_0  \sin(\omega t)} \simeq t_1 J_n(A_0)$, where $J_n(x)$ is the $n$th Bessel function of the first kind. Since $J_n(A_0)$ oscillates with the intensity ($A_0$) of the  gauge field, different limits of $t_1$ can be accessed in this experiment. Moreover, due to the high degree of controllability of most parameters in optical lattice experiments, the possibility of inducing disorder and studying its effect seems feasible after two recent  experiments \cite{billy,roati}.  

\section {Conclusion}
In conclusion, we have studied the effect of disorder in a semi-Dirac system. Using the self-consistent Born approximation, we compute the disorder-induced self-energy. Typically, for an isotropic  Dirac dispersion, the self-energy is diagonal and disorder seems to affect only the effective mass and chemical potential of the system. However, for SD systems, the self-energy is off diagonal due to the anisotropic band dispersion, which in turn leads to a topological Lifshitz transition. We furthermore chart out the different topological and trivial phases in a time-reversal and inversion symmetry broken semi-Dirac system, and show that disorder can drive transitions between all the phases. Going beyond the continuum model, we analyze these phases in a lattice model by numerically calculating the Chern number and entanglement entropy. We find that the derivative of the disorder-averaged entanglement entropy peaks near the topological transitions, which the density of states fails to signal. Thus the results obtained here due to the anisotropy in Dirac systems will hopefully inspire studying the tilted Dirac cone in borophene and variants of semi-Dirac systems with more than two Dirac nodes such as TiO$_2$/V$_2$O$_3$. The underlying lattice model for those systems may differ from the current model \cite{banerjee}, which in turn may give rise to even richer physics in the presence of disorder.  Furthermore, it has been shown \cite{song} that bond disorder behaves differently in the topological transition of a isotropic Dirac system than the site disorder studied here. Therefore it is of particular interest to study the fate of the topological states discussed here in the presence of such a disorder as well as interaction \cite{fiete2}, which we leave for future study. 
   
\section{Acknowledgements}
We thank T. Oka and T. Nag for useful discussions and suggestions.



\begin{thebibliography}{99}
\bibitem{footnote1} Note that ``semi-Dirac phase" usually refers to ideal case where the electron disperses quadratically in one direction and linearly along the orthogonal direction.
However, in this article ``semi-Dirac systems" refer to  2D Dirac systems which can host all possible distinct phases,
such as ideal single-node semi-Dirac, semi metallic with two nodes of Dirac types, and insulating semi-Dirac phase 
as a function of gap parameter, $\delta_0$ in Eq.~(\ref{ham0}).
\end{thebibliography}
\end{document}